# Superradiant coupling effects in transition-metal dichalcogenides

C. E. Stevens,[1] T. Stroucken,[2] A. V. Stier,[3] J. Paul,[1] H. Zhang,[1] P. Dey,[1] S. A. Crooker,[3] S. W. Koch,[2] AND D. Karaiskaj[1,*]

[1]Department of Physics, University of South Florida, 4202 East Fowler Ave., Tampa, Florida 33620, USA
[2]Department of Physics and Material Sciences Center, Philipps-University Marburg, Renthof 5, D-35032 Marburg, Germany
[3]National High Magnetic Field Laboratory, Los Alamos, New Mexico 87545, USA
*Corresponding author: karaiskaj@mail.usf.edu

Cooperative effects allow for fascinating characteristics in light–matter interacting systems. Here, we study naturally occurring superradiant coupling in a class of quasi–two-dimensional, layered semiconductor systems. We perform optical absorption experiments of the lowest exciton for transition-metal dichalcogenides with different numbers of atomic layers. We examine two representative materials, $MoSe_2$ and $WSe_2$, using incoherent broadband white light. The measured transmission at the A exciton resonance does not saturate for optically thick samples consisting of hundreds of atomic layers, and the transmission varies nonmonotonously with the layer number. A self-consistent microscopic calculation reproduces the experimental observations, clearly identifying superradiant coupling effects as the origin of this unexpected behavior. © 2018 Optical Society of America under the terms of the OSA Open Access Publishing Agreement

*OCIS codes:* (140.3490) Lasers, distributed-feedback; (060.2420) Fibers, polarization-maintaining; (060.3735) Fiber Bragg gratings.

## 1. INTRODUCTION

With the ability to fabricate them as monolayers, the electronic and optical properties of naturally occurring transition-metal dichalcogenide (TMDC) multilayer structures have attracted renewed interest. While excitons in monolayers are naturally confined within the layer, systematic studies of the band structure as a function of the number of layers show that even though the system becomes energetically indirect, the direct gap at the $K$-points of the Brillouin zone is preserved even in the bulk limit [1]. The near–$K$-point bands have a flat dispersion along the $K - H$ axis, indicating that even in a multilayer structure, the corresponding quasi-particles are strongly confined within the individual layers and can be considered as effectively two-dimensional. Indeed, recent theoretical and experimental studies attribute the excitonic series in TMDC multilayer systems as being due to quasi two-dimensional intra- and interlayer excitons [2,3].

Up to now, most theoretical and experimental investigations of TMDC structures have focused on the evolution of the bandgap and exciton binding, while changing the dielectric substrate or number of layers. Surprisingly, little or no attention has been paid to changes in the optical spectra of TMDC multilayer structures due to light propagation effects and radiative coupling. Therefore, the studies in this paper will focus on those specific points.

In 1954, Dicke [4] first considered the subtle situation in which two emitters decay in close proximity to each other and affect each other's emission and absorption processes, leading to cooperative behavior. In general, radiative coupling occurs if an ensemble of oscillators interacts with a common light field. The interaction of a single oscillator with the light field (re-)emitted by all oscillators in the ensemble leads to the formation of collective modes, which describe the response of the system as a whole instead of the response of the individual emitters. In fact, the emission dynamics of a single two-level system are altered by the presence of a second one, even if it is in its ground state.

As already shown by Hopfield [5], radiative coupling in a resonantly excited homogeneous semiconductor leads to the formation of exciton-polaritons with infinite radiative lifetimes. The infinite radiative lifetime is an immediate consequence of the infinite system dimensions, inhibiting an emitted photon from escaping from the sample. Instead, it is trapped in an endless cycle of (re-)absorption and (re-)emission processes, forming the hybrid exciton-light state, i.e., the exciton-polariton. Nevertheless, the polariton is damped due to nonradiative dephasing processes in its material part that lead to an irreversible (true) absorption of energy from the optical field. The concomitantly reduced light transmission can be described by the Beer–Lambert law, predicting an exponential decay of the transmitted light as a function of the optical thickness.

In contrast, a photon emitted by an individual localized oscillator escapes from the sample without re-absorption, thus opening a radiative decay-channel. The oscillator strength, $f_\lambda$, determines both, the reversible light absorption and the emission, and for an oscillator with transition energy $E_\lambda$, the associated radiative decay rate is $\Gamma_\lambda = f_\lambda E_\lambda / \hbar$ [6–9]. In an ensemble of such localized oscillators, cooperative effects depend strongly on the relative phases of the light (re-)emitted by different oscillators. If $N$ identical excited oscillators are confined to a region small compared to the optical wavelength, the overall emission is in phase, leading to a peak intensity of the emitted light field $\propto N^2$. This superlinear intensity increase is accompanied by a corresponding reduction of the radiative lifetime such that the total emitted energy, $\int I(t) dt$, increases linearly with the number of oscillators, in agreement with the requirement of energy conservation. This phenomenon is denoted in the literature as "superfluorescence" or "superradiance" [4]. For the ensemble of identical localized oscillators, the superradiant mode has the features of a single oscillator mode with $N$ times enhanced coupling strength, i.e., with $N$ times reduced radiative lifetime, and is the only optically active mode, whereas all other $(N - 1)$ modes are dark.

In solid state physics, superradiant coupling effects have been studied intensively in semiconductor multiple quantum well (MQW) structures [6–10]. In an MQW, excitons are confined to the quantum-well region, and the lateral spacing between the wells can be controlled within nanometer accuracy. The analogous situation to what was originally considered by Dicke is then realized in a so-called Bragg geometry where the center-to-center distance of neighboring quantum wells (QW) corresponds to an integer multiple of half the optical wavelength, leading to phase coherence between the emitters in the different QWs. Indeed, four-wave mixing experiments on MQW Bragg structures have demonstrated the hallmark of superradiance, a radiative decay rate increasing linearly with the QW number [10].

In the present paper, we analyze the influence of radiative coupling and light propagation effects on the optical absorption of TMDCs as a function of the number of layers in the crystal. Similar to semiconductor QWs, absorption of light in the TMDCs leads to the formation of excitons that are confined reasonably well within the individual layers. However, as compared to typical QW excitons, the excitons in TMDC monolayers display huge oscillator strengths, leading to as much as ∼10% extinction of the incoming light for a typical monoatomic layer of e.g., $WSe_2$ [11]. As has been demonstrated recently, the transmittance of a single layer can be decreased even further if it is encapsulated, e.g., by hexagonal boron nitride [12,13]. Such an encapsulation protects the TMDC monolayer from surface damage and leads to a considerable reduction of the nonradiative excitonic linewidth, manifesting itself in an observed reflection coefficient as large as 80% for a monolayer $MoS_2$. According to the Beer–Lambert law, this strong single-layer extinction should lead to a rapid decay of the transmitted light at the excitonic resonance frequency with an increasing number of atomic layers. In fact, a simple calculation of the optical density indicates that the crystal should be nearly opaque already for a few tens of atomic layers. However, the close proximity of different layers, together with the very large excitonic oscillator strength, provides favorable conditions for the occurrence of superradiant radiative coupling effects.

Whereas superradiant light emission cannot be expected in an optically indirect semiconductor, the absorption characteristics at the direct gaps at the $K$-points in a multilayer TMDC flake can still experience a superradiant coupling with a correspondingly reduced radiative lifetime. In an absorption measurement, the increased radiative decay rate then leads to a reduced percentage of the incoming energy that is irreversibly absorbed by the sample, i.e., to a reduction of the true absorption. To investigate this behavior, we perform a series of conceptually simple broadband white light transmission and reflection experiments to study the cooperative radiative coupling effects in TMDCs as a function of crystal thickness.

## 2. EXPERIMENT

Using the experimental setup shown schematically in Fig. 1, we carefully measured the optical transmission and reflection of $MoSe_2$ and $WSe_2$ samples of different thicknesses at 5 $K$. We have collected atomic force microscopy (AFM) images of all the samples studied. The thicknesses of the samples and the resulting numbers of atomic layers were obtained using AFM for each sample. In addition, for thinner crystals, Raman scattering and photoluminescence spectroscopy were used. Transmission and reflection spectra were measured quantitatively with respect to the total incoming light, and all the possible losses were carefully taken into account. The true absorption was obtained as $A = 1 - T - R = 1 - I_T/I_0 - I_R/I_0$, where $I_0$ is the intensity of the incoming signal, $I_T$ is the intensity of the transmitted signal, and $I_R$ is the intensity of the reflected signal.

The experimental transmitted, reflected, and absorption spectra are shown in Supplement 1, together with the corresponding theoretically obtained spectra. The deduced optical density data at the peak of the A exciton resonance are summarized in Fig. 2 as a function of the number of layers. We notice remarkable deviations of the measured absorption from the Beer–Lambert law, which predicts that at the A exciton resonance frequency, the sample should be essentially opaque for thicker crystals. Instead, we observe an optical density well below saturation. This phenomenon occurs in both studied systems, $MoSe_2$ and $WSe_2$, indicating that it is not unique to a specific material, but likely characteristic for the entire class of TMDC materials.

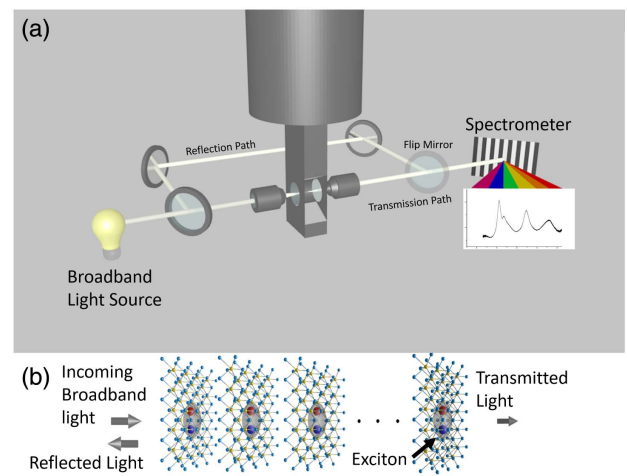

**Fig. 1.** (a) Experimental setup. The exfoliated samples are held at 5 K inside a cryostat designed for simultaneous measurement of the transmitted and reflected light. The broadband white light is focused on the sample via a 50 × microscope objective and collected on the other side by a second microscope objective. The reflected and transmitted light are dispersed by the spectrometer and collected by the CCD detector. A flip mirror is used to switch between the transmitted and reflected light paths. (b) Schematic of the light propagation through the samples.

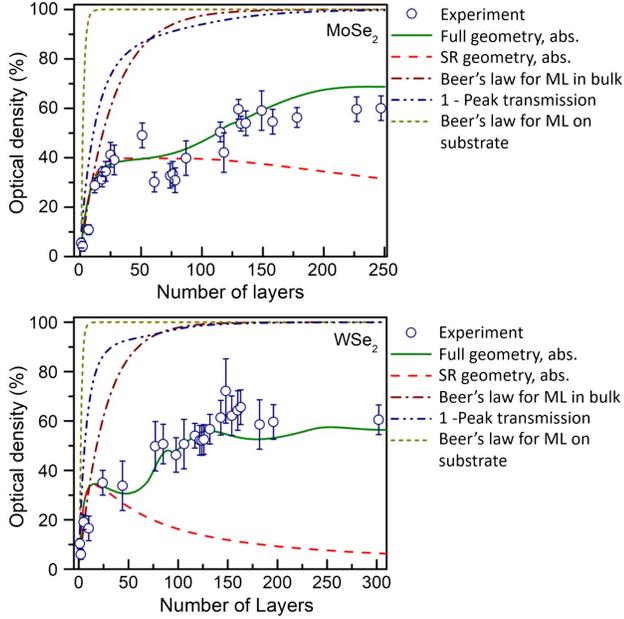

**Fig. 2.** Quantitative optical density at the exciton absorption energy for (Top) MoSe$_2$ and (Bottom) WSe$_2$ as a function of the number of layers. The blue circles are the measured optical density values. The green solid lines show the theoretically computed peak absorptions for the full geometry, and the red dashed lines show the absorption in a superradiant (SR) geometry where the interlayer spacing has been artificially set to zero. The blue dashed-dotted lines show the extinction of the transmitted intensity (1-peak transmission). For comparison, we also show the expected extinction of the transmitted intensity assuming Beer's law, once based on monolayer extinction on a quartz substrate (short-dashed) and once based on the extinction of a monolayer embedded in a dielectric with the dielectric constant of the parent bulk material. Radiative coupling is needed in the theoretical calculations (green solid line) in order to reproduce the experimentally observed light propagation as a function of layer thickness, which deviates drastically from the curves calculated using Beer's law (dark yellow and brown dashed lines).

In order to measure the lateral extent of the excitonic wave function, we measured the diamagnetic frequency shift of the A exciton under extremely high magnetic fields. For this purpose, we utilized the 65 T pulsed magnet at the National High Magnetic Field Laboratory at Los Alamos National Laboratory. Magneto-reflectance studies were performed with the samples at cryogenic temperatures down to 4 K using a home-built fiber-coupled optical probe. Broadband white light from a xenon lamp was coupled to the samples using a 100 μm diameter multi-mode optical fiber. The light was focused onto the sample at near-normal incidence using a single aspheric lens, and the reflected light was refocused by the same lens into a 600 μm diameter collection fiber. The collected light was dispersed in a 300 mm spectrometer and was detected with a liquid nitrogen cooled charge-coupled device (CCD) detector [14]. The use of very large magnetic fields allows for the observation of the small quadratic diamagnetic shift of the excitons in the TMD materials. From the diamagnetic shift, we deduce the lateral extent of the excitonic wave function to be 3 to 5 times larger than the interlayer distance. We further estimate the excitonic wave function to be well contained within the individual layers, leading to negligible surface effects for thicker samples, supported by measurements and calculations on MoS$_2$, WS$_2$, and MoTe$_2$ [2,15]. Theoretical calculations on bulk MoTe$_2$ estimate >90% of the excitonic wave function to be contained within the individual layers [2].

## 3. MICROSCOPIC THEORY

To analyze the observed phenomena, we employ a semiclassical theory of radiatively coupled two-dimensional excitons based on the semiconductor Maxwell–Bloch equations for a classical optical field interacting with equidistantly spaced two-dimensional layers. Treating the Hamiltonian of the individual layers within an effective four-band model, screening and radiative coupling of the bands under consideration is included dynamically, whereas contributions of all the other bands and the dielectric environment are contained in a background dielectric tensor [16]. For the dielectric displacement field of the layered material, we make the ansatz

$$\boldsymbol{D} = \epsilon_\parallel \boldsymbol{E}_\parallel + \epsilon_\perp E_z e_z + 4\pi \boldsymbol{P}, \quad (1)$$

where $\epsilon_\parallel$ and $\epsilon_\perp$ represent the in-plane and out-of-plane background dielectric constants, respectively, and $\boldsymbol{P}$ describes the nonlocal, time- and frequency-dependent resonant contributions of the bands under consideration. (We use c. g. s. units throughout the theory.)

Using the generalized Coulomb gauge, $\epsilon_\parallel \nabla_\parallel \cdot \boldsymbol{A}_\parallel + \epsilon_\perp \partial_z A_z = 0$, and restricting the analysis to light propagation along the $c$ axis, a division into transverse and longitudinal contributions yields the wave equation for the transverse optical field and Poisson's equation for the scalar potential:

$$\left(-\partial_z^2 - \epsilon_\parallel(z)\frac{\omega^2}{c^2}\right)\boldsymbol{E}_\parallel^T(z,\omega) = 4\pi\frac{\omega^2}{c^2}\boldsymbol{P}_\parallel^T(\omega), \quad (2)$$

$$(-\epsilon_\perp \partial_z^2 + \epsilon_\parallel \boldsymbol{q}_\parallel^2)\phi(\boldsymbol{q}_\parallel, z) = 4\pi(\rho - \nabla \cdot \boldsymbol{P}^L), \quad (3)$$

where $\boldsymbol{E}^T = -\frac{1}{c}\dot{\boldsymbol{A}}$; $\boldsymbol{E}^L = -\nabla\phi$; $\boldsymbol{A}$ and $\phi$ are the vector and scalar potentials; and the superscripts $T$ and $L$ denote transverse and longitudinal parts, respectively.

On the length scale of the optical wavelength, the resonant quasi–two-dimensional contributions of the individual layers can be approximated by

$$\boldsymbol{P} = \sum_n \boldsymbol{P}_{n\parallel}\delta(z - z_n),$$

where $z_n = (n - 1/2)d$ is the central position of the $n$th layer, $d$ is the natural layer-to-layer distance, and $\boldsymbol{P}_{n\parallel}$ is the optically induced polarization in the $n$th layer.

The radiative coupling effects result from the interaction of the individual layers with the light emitted by all layers in the sample [9]. The total optical field at the position of the $n$th layer can then be expressed as

$$\boldsymbol{E}_\parallel^T(z_n, \omega) = \boldsymbol{E}_\parallel^{\text{in}}(z_n, \omega) + 4\pi\frac{\omega^2}{c^2}\sum_m G_0(z_n, z_m, \omega)\boldsymbol{P}_{m\parallel}^T(\omega), \quad (4)$$

where $\boldsymbol{E}_\parallel^{\text{in}}(z, \omega)$ represents the incoming field with a dispersion determined by the dielectric environment of the nonresonant background and substrate, and $G_0(z, z', \omega)$ describes the radiation of a localized oscillator embedded within this environment. As the material polarization is induced by the total transverse field, the second part of Eq. (4) results in a dynamical dipole–dipole interaction mediated by the optical field.

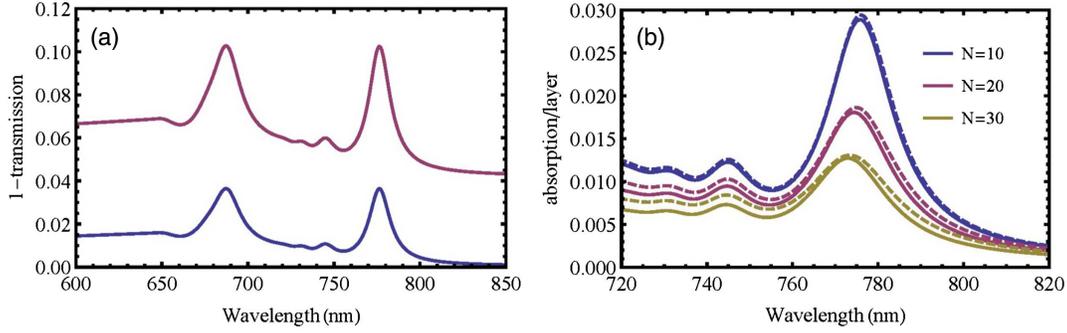

**Fig. 3.** (a) Calculated transmission of a single MoSe$_2$ layer embedded in bulk (blue curve) and of a monolayer with the same linear susceptibility on a BK7 substrate. (b) Comparison of the absorption per layer to the layer absorption in a superradiant geometry for different sample thicknesses. The solid lines show the calculated layer absorption of MoSe$_2$ on a BK7 substrate, computed for the full geometry. The dashed lines show the corresponding results in a superradiant geometry, i.e., neglecting the spacing between different layers in the wave equation. All spectra have been calculated using a nonradiative homogenous linewidth of 15 meV for the A exciton series and 25 meV for the B exciton series.

In the linear regime, the transverse part of the optically induced layer polarization is related to the transverse optical field via $P^T_{n\|}(\omega) = \chi^T_n(\omega) E^T_\|(z_n, \omega)$, where $\chi^T_n(\omega)$ is the transverse susceptibility of the $n$th layer. To compute this susceptibility, we employ the widely used massive Dirac-fermion (MDF) model [17], where the Hamiltonian describes an effective four-band system adapted to the symmetry properties of the hexagonal lattice. In reciprocal space, the first Brillouin zone has the same honeycomb geometry as in real space, and a direct gap occurs at the corners of the Brillouin zone, i.e., the $K^\pm$ Dirac points. The nonequivalent Dirac points are identified by the so-called valley index, $\tau = \pm 1$, and they are related by the parity transformation. The near–$K$-point quasi-particles are described as relativistic Dirac fermions with pseudospins that couple to the light field. In a naturally A-A′–stacked multilayer structure, adjacent layers are related by the parity transformation. Thus, the joint Brillouin zone of the multilayer structure can be considered to be composed of mirror-identical copies of the 2D monolayer Brillouin zone, leading to a layer-dependent valley index, $\tau_n = \pm(-1)^n$.

For the individual layers, we compute the layer susceptibility microscopically from the coupled gap equations and the Dirac–Bloch equations (DBE), as described in detail in Ref. [16]. Here, we have a single free parameter, which can be related to the physical thickness of a single monolayer. All other parameters are obtained from density functional theory (DFT) calculations. Since the coupled equations contain the screened Coulomb interaction, which is obtained as solution of Poisson's Eq. (3) for the electron and hole charge density, $\rho$, they contain a Coulomb-mediated interlayer interaction, allowing for a seamless description of the transition from monolayer to bulk. For the numerical evaluations, we use the material parameters listed in Ref. [16].

As screening by the surrounding layers influences the spectral position of the exciton and its oscillator strength, the different dielectric environments of the individual layers lead to slightly different resonance frequencies. Therefore, the response of a multilayer structure is generally inhomogeneously broadened due to the individual contributions from the different layers. However, this effect becomes negligibly small for sufficiently thick structures where the distance of most layers from the sample surfaces well exceeds the in-plane exciton Bohr radius. In such cases, the linear susceptibilities of the individual layers within the sample become identical, influenced only by intrinsic screening effects.

Alltogether, the transmission and absorption coefficients of a multilayer sample reflect the interplay between the individual layer susceptibilities, the dielectric environment, and the propagation effects. For a monolayer embedded between two dielectrics with refractive indices $n_T = \sqrt{\epsilon^T_\|}$ and $n_B = \sqrt{\epsilon^B_\|}$, the amplitude transmission, $t(\omega)$, and reflection, $r(\omega)$, coefficients relating the transmitted and reflected fields to the incoming field are given by

$$t(\omega) = \frac{n_T}{(n_T + n_B)/2 - 2\pi i \frac{\omega}{c} \chi(\omega)},$$

$$r(\omega) = t(\omega) - 1.$$

The transmitted and reflected intensities are related to the amplitude coefficients via $I_T = n_B |tE_0|^2$, $I_R = n_T |rE_0|^2$, and the incoming intensity is given by $I_0 = n_T |E_0|^2$. One thus finds for the true absorption coefficient of a monolayer,

$$\alpha(\omega) = \frac{4\pi n_T \frac{\omega}{c} \mathrm{Im}[\chi^T(\omega)]}{|(n_T + n_B)/2 - 2\pi i \frac{\omega}{c} \chi^T(\omega)|^2}.$$

For a single 2D oscillator with resonance frequency $\omega_0$ and nonradiative decay rate $\gamma$, one finds a true absorption coefficient: $\alpha(\omega) = \frac{n_T}{\bar{n}} \gamma \Gamma/\bar{n}/((\omega - \omega_0)^2 + (\gamma + \Gamma/\bar{n})^2)$, where $\Gamma$ is the radiative decay rate for the monolayer in vacuum, and $\bar{n} = (n_T + n_B)/2$ is the average refractive index of the top and bottom materials. As a true absorption process not only requires an initial excitation but also a subsequent scattering into an optically dark state, the peak absorption is determined by the interplay between radiative and nonradiative decay rates and reaches its maximum value if both are equal. For a typical nonencapsulated monolayer on top of a fused silica substrate, this gives an extinction coefficient of 10% to 20% of the incoming intensity, depending on the specific material and the nonradiative homogeneous linewidth. Neglecting the changes in the dielectric environment and radiative coupling, the Beer–Lambert law would predict an exponential decay of the transmitted intensity, $T_N = \frac{n_B}{n_T} |t|^{2N}$, leading to an extinction coefficient of roughly 90% for a sample consisting of only five layers, in strong disagreement with the experimentally observed opacity.

An increase in the number of layers modifies the transmission and absorption characteristics in two ways: first, it changes the effective background refractive index experienced by the individual layers, and, second, it leads to radiative coupling. The effect of changing the dielectric environment alone is best illustrated by comparing the transmitted intensity of a monolayer on a substrate with that of a monolayer with the same linear susceptibility embedded in bulk. Such a comparison is shown in Fig. 3(a) for the example of $MoSe_2$, using the bulk layer susceptibility. For the background in-plane dielectric constant of $MoSe_2$ we use $\epsilon_\parallel = 5.01$, which has been extracted by combining bulk DFT calculations, with the analytically computed contributions, with the longitudinal in-plane dielectric constant [16]. As can be recognized, the increased background refractive index in the bulk environment reduces the layer extinction by roughly a factor of 2.5. The predictions for the transmitted intensity based on Beer's law, once using the monolayer-on-substrate transmission coefficient and once using the monolayer-in-bulk transmission coefficient, are shown as dark yellow dashed and brown dashed-dotted curves in Fig. 2, respectively. As can be recognized, the increased background refractive index of the bulk environment decreases the opacity of a multilayer sample predicted by Beer's law considerably, yielding saturation only after roughly 100 layers.

Apart from simple dielectric effects, the optical spectra are modified by radiative coupling features, which depend sensitively on the phase coherence between the emitters in different layers. Moderately increasing the number of layers and ensuring that the total sample thickness, $N \times d$, remains much smaller than the optical wavelength in the material, all emitters/absorbers are in phase and the emitted fields from the different layers interfere constructively. Here, the superradiant mode is the only bright one, and the system effectively behaves as a single emitter with an $N$ times enhanced coupling strength.

Since the radiative lifetime is inversely proportional to the coupling strength, the enhanced coupling in the superradiant geometry reduces the radiative lifetime and increases the radiative broadening. While the effect of the modified dielectric environment is clearly observable in the transmission characteristics, the radiative broadening is best observed in the true absorption spectra. In Fig. 3(b), the dashed lines show the absorption per layer for a sample containing 10, 20, and 30 layers in a superradiant geometry on a BK7 glass substrate. As we can see, the superradiant coupling reduces the absorption per layer with an increasing number of layers, with a correspondingly increased linewidth (FWHM). The full lines in Fig. 3(b) show the theoretically predicted absorption/layer computed for the corresponding samples on a BK7 substrate using the full geometry. The strong similarity of the spectra shows that, within this thickness range, the radiative coupling is essentially superradiant.

If the total sample thickness is further increased such that it exceeds the optical wavelength, the system essentially develops into an effectively homogeneous medium with a polariton dispersion

$$k_z^2 = \frac{\omega^2}{c^2}(\epsilon_\parallel + 4\pi\chi^T(\omega)/d),$$

giving Beer's law for the transmitted intensity of an incoming plane wave:

$$I(\omega, z) = I_0(\omega)e^{-\alpha(\omega)z}$$

with an absorption coefficient $\alpha(\omega) \approx \frac{4\pi\omega}{n(\omega)c}\frac{\chi''(\omega)}{d}$ and refractive index $n(\omega) = \text{Re}[\sqrt{\epsilon_\parallel + 4\pi\chi^T/d}]$, determined by the 2D optical susceptibility.

## 4. THEORY–EXPERIMENT COMPARISON

In order to test our model assumptions, we probe the lateral extent of the excitonic wave function by measuring the diamagnetic frequency shift of the A exciton under very high magnetic fields. The energy position of the A exciton resonance as a function of energy for magnetic fields up to 60 T perpendicular to the layers or parallel to the $c$ axis is shown in Figs. 4(a) and 4(c) for $WSe_2$ and $MoSe_2$, respectively. From the energy position of the A exciton at +60 T and −60 T, we obtain the diamagnetic shift shown in Figs. 4(b) and 4(d) for $WSe_2$ and $MoSe_2$, respectively.

The diamagnetic shift is fitted using a simple quadratic formula, which yields shifts of the average position of $\sim 0.77$ µeV/T$^2$ and $\sim 0.12$ µeV/T$^2$ for $WSe_2$ and $MoSe_2$, respectively. The lateral exciton size is obtained from the diamagnetic shift via $r_X = \sqrt{\langle r \rangle^2} = \sqrt{8m_r\sigma}/e$, where $m_r$ is the exciton reduced mass. Using $m_r/m_0 = 0.17$ for $WSe_2$ and $m_r/m_0 = 0.27$ for $MoSe_2$, we obtain $r_X = 2.44$ nm and $r_X = 1.21$ nm for $WSe_2$ and $MoSe_2$, respectively. Using the natural layer-to-layer distance of $d = 6.5$ Å for both materials, the exciton radius is roughly 2 to 4 times larger than the interlayer distance, resulting in negligible surface effects for samples with $N \gtrsim 7$.

To analyze the experimental results, we use the values $\epsilon_\parallel = 5.01$ and $\epsilon_\parallel = 13.6$ for the bulk in-plane dielectric constants for $MoSe_2$ and $WSe_2$, respectively. These are determined by subtracting the resonant contributions to the dielectric constant from those obtained from DFT calculations for the respective bulk parent materials [18] and from the measured values at 1024 nm [19], where the resonant contributions have been computed analytically within the MDF model [16]. To avoid rapid oscillations from the substrate/vacuum interface that cannot be resolved experimentally, we perform simulations for a half-space geometry, i.e., air/sample/substrate. Furthermore, we use a phenomenological nonradiative linewidth of 15 meV for the A- and 25 meV for the B-excitonic series for *all* samples. Though the assumption of equal nonradiative decay rates in different samples is questionable in principle, this is a reasonable assumption for the larger samples where surface effects become negligible, and it allows us to identify changes in the absorption characteristics due to radiative coupling effects only. The resulting calculated peak absorption and transmission as a function of the number of layers (sample thickness) is plotted in Fig. 2 for $MoSe_2$ (Top) and $WSe_2$ (Bottom) as green and blue-dashed lines, respectively. In particular, the true peak absorption (green lines) exhibits a behavior that differs significantly from the calculated absorption following the Beer–Lambert law, be it based on the monolayer-on-substrate transmission (dark yellow dashed line) or based on the monolayer-in-bulk transmission (brown dashed-dotted line). As we can see, the experimental data are reasonably well reproduced by the calculated true absorption. Furthermore, we experimentally observe a weak, nearly constant absorption of the multilayer structure over the range $N = 15$ to $N \sim 50$, in stark contradiction to the exponential decay of the transmitted intensity prescribed by the Beer–Lambert law.

Under superradiant coupling conditions, the peak absorption at resonance is proportional to $\gamma N\Gamma/(\gamma + N\Gamma)^2$, where $\gamma$ and $\Gamma$ are the nonradiative and radiative decay rates, respectively. In Fig. 2, the superradiant peak absorption is plotted as a red dashed line. For $MoSe_2$, the numerical evaluations give a radiative decay rate of $\hbar\Gamma = 0.6$ meV. Estimating the nonradiative linewidth to be 15 meV, the maximum peak absorption is expected for $N \approx 20$, which decays slowly in the region between $N = 20$ and $N = 50$. These features are characteristic for superradiant coupling and in good agreement with the experimental observations. For samples containing more than roughly 60 layers, we observe deviations from the superradiant coupling. Here, i.e., at a total sample thickness of roughly a quarter-wavelength, the system transitions from the superradiant to the bulk regime.

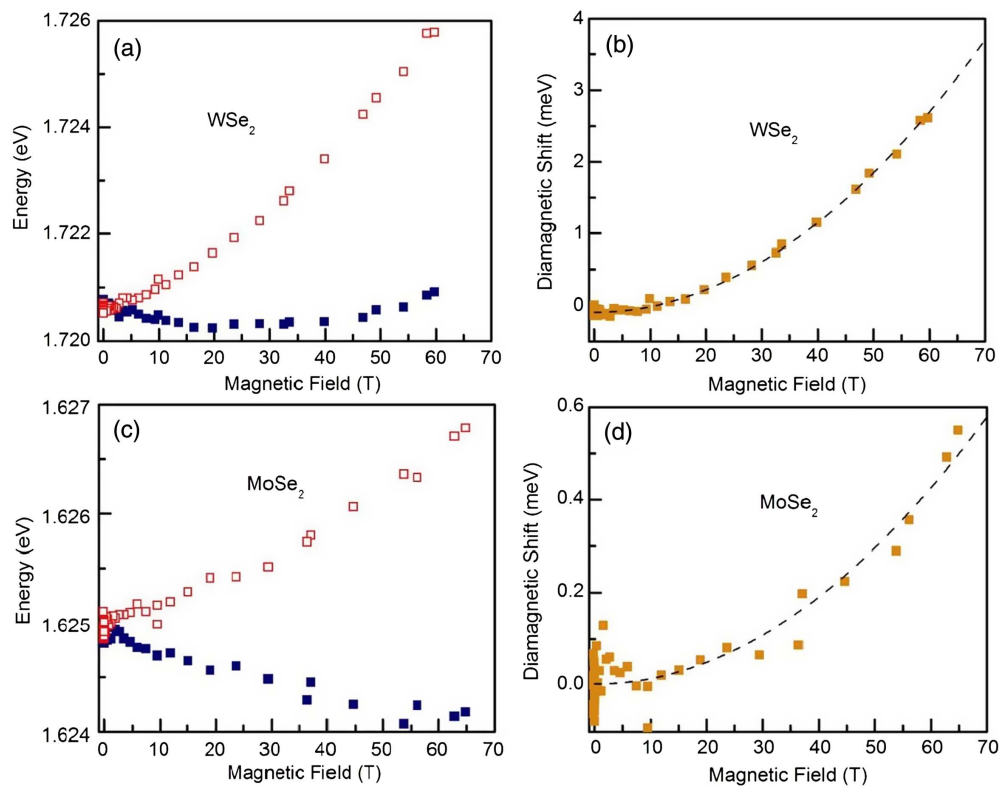

**Fig. 4.** (a) Energy position of the A exciton in WSe$_2$ as a function of magnetic field strength up +60 T (blue squares) and up to −60 T esla (red squares). (b) Obtained diamagnetic shift for the A exciton in WSe$_2$ (yellow squares). The dashed line is a quadratic fitting. (c) Energy position of the A exciton in MoSe$_2$ as a function of magnetic field strength up +60 T (blue squares) and up to −60 T (red squares). (d) Obtained diamagnetic shift for the A exciton in MoSe$_2$ (yellow squares). The dashed line is a quadratic fitting.

Due to the larger background in-plane dielectric constant, the optical wavelength in the WSe$_2$ sample at the resonance frequency is only roughly 120 nm. Correspondingly, the transition from the superradiant regime to the bulk limit occurs for smaller sample sizes with $N \approx 45$, which is again in good agreement with the experimental observations.

In Fig. 5, we plot the full absorption spectra measured experimentally and calculated theoretically for both MoSe$_2$ and WSe$_2$

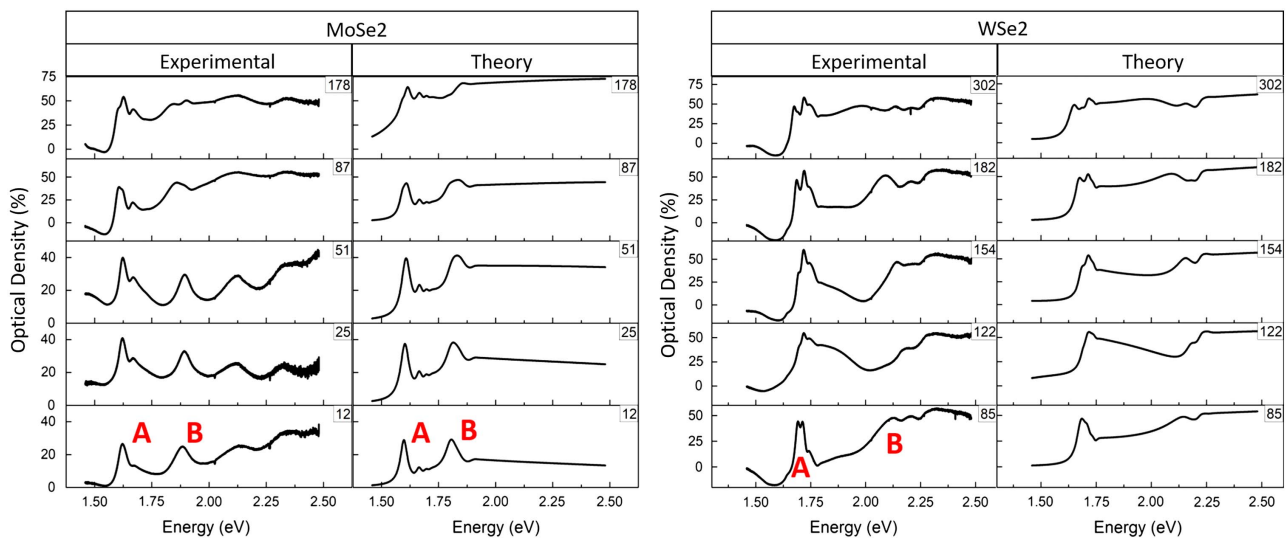

**Fig. 5.** Experimental and calculated absorption spectra for (Left) MoSe$_2$ and (Right) WSe$_2$ for different sample thicknesses (number of layers).

for different numbers of layers. We obtain quantitative agreement between experiment and theory in terms of the absolute absorption strength. The experimental line shapes are well reproduced by the theoretical calculations that include superradiant coupling, which further reaffirms the validity of the theoretical approach and the superradiant nature of the light propagation through layered TMDCs.

## 5. SUMMARY AND CONCLUSIONS

In conclusion, we analyze the optical transmission/absorption properties of exfoliated TMDC samples for different numbers of layers, ranging all the way from monolayers to bulk-like configurations with more than a hundred layers. We carefully characterize the respective sample thicknesses using AFM; for thinner samples, photoluminescence and Raman spectroscopy were additionally used. The absorption was quantitatively measured for the A exciton resonance by carefully measuring transmission and reflection simultaneously, and by taking into account all the possible losses. The resulting optical density as a function of sample thickness deviates significantly from the Beer–Lambert law. We model the observed effects employing the semiconductor Maxwell–Bloch equations for a classical optical field interacting with equidistantly spaced two-dimensional layers, where the A excitons are assumed to be well localized within the individual layers. To validate the model assumptions, we carefully measure the excitonic Bohr radii in the two representative bulk TMDs using the diamagnetic energy shift of the excitonic resonance at magnetic fields up to 65 T. The exciton wave functions obtained from these measurements are well localized within the layers and are also in agreement with recent angle-resolved photoemission studies performed in bulk $WSe_2$, which reveal a two-dimensional character of the bands at the $K$-point [20]. The theoretical calculations reveal that the experimentally observed variation of the resonant optical absorption can be uniquely attributed to the superradiant coupling between the excitons in the different TMDC layers.

**Funding.** Basic Energy Sciences (BES) (DE-SC0012635); Deutsche Forschungsgemeinschaft (DFG) (DFG:SFB1083); Directorate for Mathematical and Physical Sciences (MPS) (NSF DMR-1157490).

See Supplement 1 for supporting content.